# Doping Effect of Nano-Diamond on Superconductivity and Flux Pinning in $MgB_2$

C.H. Cheng[1*], H. Zhang[1,2], Y. Zhao[1,2], Y. Feng[3], X.F. Rui[2], P. Munroe[1], H.M. Zeng[4], N. Koshizuka[5], M. Murakami[5]

[1]School of Materials Science and Engineering, University of New South Wales, Sydney 2052, NSW, Australia

[2]State Key Lab for Mesophysics, Department of Physics, Peking University, Beijing 100871, China

[3]Northwest Institute for Nonferrous Metal Research, P.O. Box 51, Xi'an, Shaanxi, 710016, China

[4]Key Laboratory of Polymeric Composites and Functional Materials, The Ministry of Education, Zhongshan University, Guangzhou, 510275, People's Republic of China,

[5]Superconductivity Research Laboratory, ISTEC, 1-10-13 Shinonome, Koto-ku, Tokyo, 135-0062, Japan.

Doping effect of diamond nanoparticles on the superconducting properties of $MgB_2$ bulk material has been studied. It is found that the superconducting transition temperature $T_c$ of $MgB_2$ is suppressed by the diamond-doping, however, the irreversibility field $H_{irr}$ and the critical current density $J_c$ are systematically enhanced. Microstructural analysis shows that the diamond-doped $MgB_2$ superconductor consists of tightly-packed $MgB_2$ nano-grains (~50-100 nm) with highly-dispersed and uniformly-distributed diamond nanoparticles (~10-20 nm) inside the grains. High density of dislocations and diamond nanoparticles may take the responsibility for the enhanced flux pinning in the diamond-doped $MgB_2$.

*Corresponding author: email: c.cheng@unsw.edu.au

## I Introduction

Since the discovery of superconductivity at 39 K in $MgB_2$ [1], significant progress has been made in improving the performance of $MgB_2$ materials [2-7]. $MgB_2$ offers the possibility of wide engineering applications in the temperature range 20-30 K, where conventional superconductors, such as $Nb_3Sn$ and Nb-Ti alloy, cannot play any roles due to their low $T_c$. However, the realization of large-scale applications for $MgB_2$-based superconductivity technology essentially relies on the improvement of the pinning behaviour of $MgB_2$ in high fields. As it has poor grain connection and a lack of pinning centres, $MgB_2$ often exhibits a rapid decrease in critical current density, $J_c$, in high magnetic fields. Fortunately, through the formation of nanoparticle structures in bulk $MgB_2$ [3-5] and thin films [7], the problem of the poor grain connection can be solved, and the flux pinning force can also be significantly enhanced due to an increase of pinning centres served by grain boundaries. In order to improve further the performance of $MgB_2$, it is necessary to introduce more pinning centres, especially those consisting of nano-sized second-phase inclusions which often provide strong pinning forces.

Nano-diamond, prepared by the detonation technique, has been widely used as an additive to improve the performance of various materials [8]. However, nano-diamond-doping effect on the superconducting properties $MgB_2$ has never been reported although carbon with other forms has been used as dopants in $MgB_2$ [9-11]. The high dispersibility of the nano-diamond powder makes it possible to form a high density of nano-inclusions in $MgB_2$ matrix. In this article, we have investigated the doping effect of nano-diamond on the superconducting properties of $MgB_2$. Our results show that the nano-diamond-doped $MgB_2$ consists of tightly-packed $MgB_2$ nano-grains (~50-100 nm) with diamond nanoparticles (~10-20 nm) wrapped within the grains. This unique microstructure provides the samples with a good grain connection for the $MgB_2$ phase and a high density of flux-pinning centres served by the diamond nanoparticles. Compared to the $MgB_2$ bulk materials doped with other nanoparticles [3-7], the irreversibility line has been significantly improved and the $J_c$ in high magnetic fields has been largely increased in these nano-diamond-doped $MgB_2$.

## II. Experimental

The $MgB_2$-diamond nanocomposites with compositions of $MgB_{2-x}C_x$ (x=0, 5%, 8%, and 10%) were prepared by solid-state reaction at ambient pressure. Mg powder (99% purity, 325 meshes), amorphous B powder (99% purity, submicron-size), and nano-diamond powder (10-20 nm) were mixed and ground in air for 1 h. An extra 2% of Mg powder was added in the starting materials to compensate the loss of Mg caused by high temperature evaporation. The mixed powders were pressed into pellets with dimensions of 20x10x3 $mm^3$ under a pressure of 800 $kg/cm^2$, sandwiched into two MgO plates, sintered in flowing Ar at 800 $^oC$ for 2 h, and then quenched to room-temperature in air. In order to compare the substitution effect of carbon in boron in $MgB_2$ with the additional effect of the nano-diamond in $MgB_2$, a sample with an added 1.5 wt% of nano-diamond in $MgB_2$ was prepared. The sintering temperature and the sintering time for this sample were reduced respectively to 730 $^oC$ and 30 min in order to reduce the chemical reaction between the $MgB_2$ and the diamond. This sample has been referred to as "1.5wt%C".

The crystal structure was investigated by powder x-ray diffraction (XRD) using an X'pert MRD diffractometer with Cu $K\alpha$ radiation. The microstructure was analysed with a Philips CM200 field emission gun transmission electron microscope (FEGTEM). DC magnetization measurements were performed in a superconducting quantum interference device (SQUID, Quantum Design MPMS-7). $J_c$ values were deduced from hysteresis loops using the Bean model. The values of the irreversibility field, $H_{irr}$, were determined from the closure of hysteresis loops with a criterion of $10^2$ $A/cm^2$.

## III. Results and Discussion

Figure 1 shows the XRD patterns of the nano-diamond powder and the typical $MgB_2$-diamond composites. The reflection (111) of the diamond is extremely broad and an amorphous-phase-like-background can be seen in the XRD pattern. The particle size of the nano-diamond powder is estimated to be about 20 nm according to the width of the reflection. In relation to the nano-diamond-doped $MgB_2$, one of the impurity phases is MgO, which may have formed during the mixing of raw materials in air. Diamond should be present as another impurity phase in the composites; however, its main reflection (111) cannot be seen in XRD patterns, due to an overlap with the $MgB_2$ (101) peak. With increasing doping level, an amorphous-phase-like background in the XRD pattern gradually appears, suggesting the existence of unreacted nano-diamond in the sample. As for the diamond-added $MgB_2$ sample (1.5wt%C), which contains an x=5.4% equivalent percentage of carbon atoms, the background of its XRD pattern shows some similarity to the background of the nano-diamond, suggesting that a substantial amount of unreacted nano-diamond exists within this sample.

Figure 2 shows the doping level dependence of the lattice constants for the nano-diamond-doped $MgB_2$. The length of *c*-axis does not exhibit significant change with increasing nano-diamond doping level, but the *a*-axis is systematically decreased by doping the nano-diamond, indicating that a certain amount of carbon atoms have substituted for boron atoms in $MgB_2$. This result is consistent with those reported by other groups, which show that partial substitution of boron by carbon results in a decrease of the lattice parameter [9-11].

Figure3 shows the temperature dependence of magnetization at 2 mT for nano-diamond-doped $MgB_2$. The superconducting transition temperature $T_c$ decreases with increasing doping level. This result is consistent with those reported by other group on the carbon-doped $MgB_2$, which suggest that carbon can partially substitute boron in $MgB_2$, decrease the carrier (hole) concentration, and consequently decrease the $T_c$ [9-10]. The doping level dependence of $T_c$ is shown in the inset of Fig.3. The values of onset $T_c$ for these carbon-substituted $MgB_2$ samples are 38.6 K for x=0, 36.1 K for x=5%, 33.0 K for x=8%, and 31.3 K for x=10% (see the open circles). The $T_c$ for the sample 1.5wt%C is 36.9 K (see the closed circle), which is higher than that for the sample of x=5% ($T_c$=36.1 K), despite the former having a higher equivalent atomic percentage of carbon (x=5.4%). This result suggests that only a part of carbon have been doped into the crystal structure of $MgB_2$ in the diamond-added sample, consistent with the XRD analysis.

Figure 4 shows the magnetic field dependence of $J_c$ at 10, 20, and 30 K for the carbon-substituted $MgB_2$ samples. At 30 K, the undoped $MgB_2$ exhibits the highest $J_c$ and the slowest decrease of $J_c$ with $H$; whereas the sample of x=10% shows the lowest $J_c$ and the quickest drop of $J_c$ with $H$. It is evident that the $J_c$-$H$ behaviour at 30 K for these samples is positively correlated to their $T_c$ values. However, when the temperature decreases to the values far below $T_c$, a totally different situation appears. For example, at 10 K and 20 K, the diamond-doped samples show a much better $J_c$-$H$ behaviour. The $J_c$ drops much more slowly in diamond-doped samples than in pure $MgB_2$. The best $J_c$ at 20 K is found in the sample of x=10%, reaching a value of $6x10^3$ A/cm$^2$ in a 4 T field, indicating that a strong flux pinning force exists in these diamond-doped samples.

However, the effect of diamond doping on the enhancement of flux pinning in $MgB_2$ may be counterbalanced by its suppression on superconductivity, as clearly shown in the situation of $T$=30 K (see Fig.3). This counterbalancing effect may also exist at other temperatures, even when the effect of the $J_c$-enhancement is dominant. The further increase of $J_c$ depends critically on reducing the $T_c$-suppression effect in the $MgB_2$-diamond composite. This idea is confirmed by the results obtained in the diamond-added sample, 1.5wt%C, which has a higher $T_c$ than other diamond-doped samples (see Fig.3) and contains more nano-diamond inclusions as suggested by the XRD analysis (see Fig.1) and confirmed by our TEM analysis shown below. As shown in the inset of Fig. 4, the diamond-added sample shows a much better $J_c$-$H$ behaviour than the carbon-substituted sample. Its $J_c$ reaches $1x10^4$ $A/cm^2$ at 20 K and 4 T, and its $H_{irr}$ reaches 6.4 T at 20 K. In fact, at all temperatures below 35 K, the $J_c$-$H$ behaviour (results at 20 K are shown here only) of the diamond-added sample are much better than those of other samples in this study.

The $H_{irr}$-$T$ relations for the diamond-doped $MgB_2$ are shown in figure 5. The $H_{irr}$ ($T$) curves get steeper with increasing doping level, although the substitution of carbon for boron decreases the superconducting transition temperature. The best value of $H_{irr}$ for these diamond-substituted $MgB_2$ samples including x=0, 5%, 8%, and 10% is found in the sample with x=10%, reaching 5.7 T at 20 K. However, the diamond-added $MgB_2$ (1.5wt%C) shows a better $H_{irr}$-$T$ behaviour than the diamond-substituted $MgB_2$. The $H_{irr}$ value reaches 6.3 T at 20 K. The result clearly shows that the diamond doping does enhance the flux pinning in $MgB_2$ significantly, and also suggests that the pinning behaviour of the nano-diamond-doped $MgB_2$ depends on the density of the unreacted diamond nanoparticles in the samples.

. Figure 6 shows the typical results from microstructural analysis for the diamond-substituted $MgB_2$ (Fig.6a) and diamond-added $MgB_2$ samples (Fig.6b). The diamond-substitutional sample mainly consists of relatively large $MgB_2$ grains (~1 micron or so in size) with a high density of dislocations. In some areas, discrete nano-sized particles can be seen. The diamond-added sample mainly consists of two kinds of nanoparticles: $MgB_2$ grains with a size of 50-100 nm and diamond particles with a size of 10-20 nm. In fact, this diamond-added $MgB_2$ forms a typical nanocomposite material. The nano-diamond particles are inserted into the $MgB_2$ grains. As the *ab*-plane coherence length of $MgB_2$ is about 6-7 nm [12], these 10- to 20-nm-sized diamond inclusions, with a high density, are ideal flux pinning centres and are responsible for the high performance in our samples.

It is worth noting that the enhancement of flux pinning in nano-diamond-doped $MgB_2$ is even better than the Ti-doped $MgB_2$ [3-5], where the $TiB_2$ nanoparticles mainly stay in grain boundaries. This suggests that a highly dispersed distribution of nano-inclusions in $MgB_2$ is more effective in enhancing flux pinning than a more localised distribution in the grain boundaries. Besides, compared to the $Y_2O_3$ nano-particle doping, an advantage of the nano-diamond doping is that the lattice contact of the cubic diamond ($a$=0.356 nm) is very close to the $c$-axis of $MgB_2$ ($c$=0.352 nm). Therefore, these diamond nanoparticles may provide nucleation centres for $MgB_2$ and are tightly bound to them. This may explain why our $MgB_2$-diamond nanocomposite performs much better than the nano-$Y_2O_3$-doped $MgB_2$ [6]. It is expected that the performance of the $MgB_2$-diamond nanocomposite may be further improved by optimising the microstructure and the doping levels.

## IV. Summary and Conclusions

In summary, we have investigated the doping effect of diamond nanoparticles on the superconducting properties of $MgB_2$ bulk material. We have observed that the superconducting transition temperature $T_c$ of $MgB_2$ is suppressed by the diamond-doping, like doping with other forms of carbon. However, the irreversibility field $H_{irr}$ and the critical current density $J_c$ are systematically enhanced by doping nano-diamond. Microstructural analysis shows that the diamond-doped $MgB_2$ superconductor consists of tightly-packed $MgB_2$ nano-grains (~50-100 nm) with highly-dispersed and uniformly-distributed diamond nanoparticles (~10-20 nm) inside the grains. High density of dislocations and diamond nanoparticles may be responsible for the enhancement of the flux pinning in the nano-diamond-doped $MgB_2$.

**Acknowledgement** The authors are grateful to Miss Sisi Zhao for her helpful discussion in preparing the manuscript. This work was supported in part by the University of New South Wales through the Goldstar Award for Cheng. Financial support from the Ministry of Science and Technology of China (NKBRSF-G19990646) is also acknowledged.

**Figure Captions:**

Figure 1 Powder XRD patterns for nano-diamond-doped $MgB_2$. The pattern on the top row is for the nano-diamond.

Figure 2 Doping dependence of lattice constants for nano-diamond-doped $MgB_2$.

Figure 3 Temperature dependence of magnetization at 2 mT for nano-diamond-doped $MgB_2$. Inset: Doping level dependence of $T_c$. The closed circles represent the results for the sample 1.5wt%C in both the main and the inset figures.

Figure 4 Magnetic field dependence of $J_c$ at 10, 20, and 30 K for nano-diamond substituted $MgB_2$ with x=0 (dashed lines), 5% (solid lines), 8% (solid circles), and 10% (opened triangles). Inset: Comparison of $J_c$-$H$ behaviour at 20 K between the diamond-added sample (1.5wt%C) with some diamond-substituted samples (x=0, 5%, and 10%).

Figure 5 Variation of $H_{irr}$ with temperature $T$ for nano-diamond doped$MgB_2$.

Figure 6 FEGTEM micrographs for (a) diamond-substituted $MgB_2$ with x=5%; (b) diamond-added $MgB_2$ with the carbon content of 1.5 wt%. The atomic percentages of carbon in the sample 1.5wt%C is equivalent to x=5.4%.

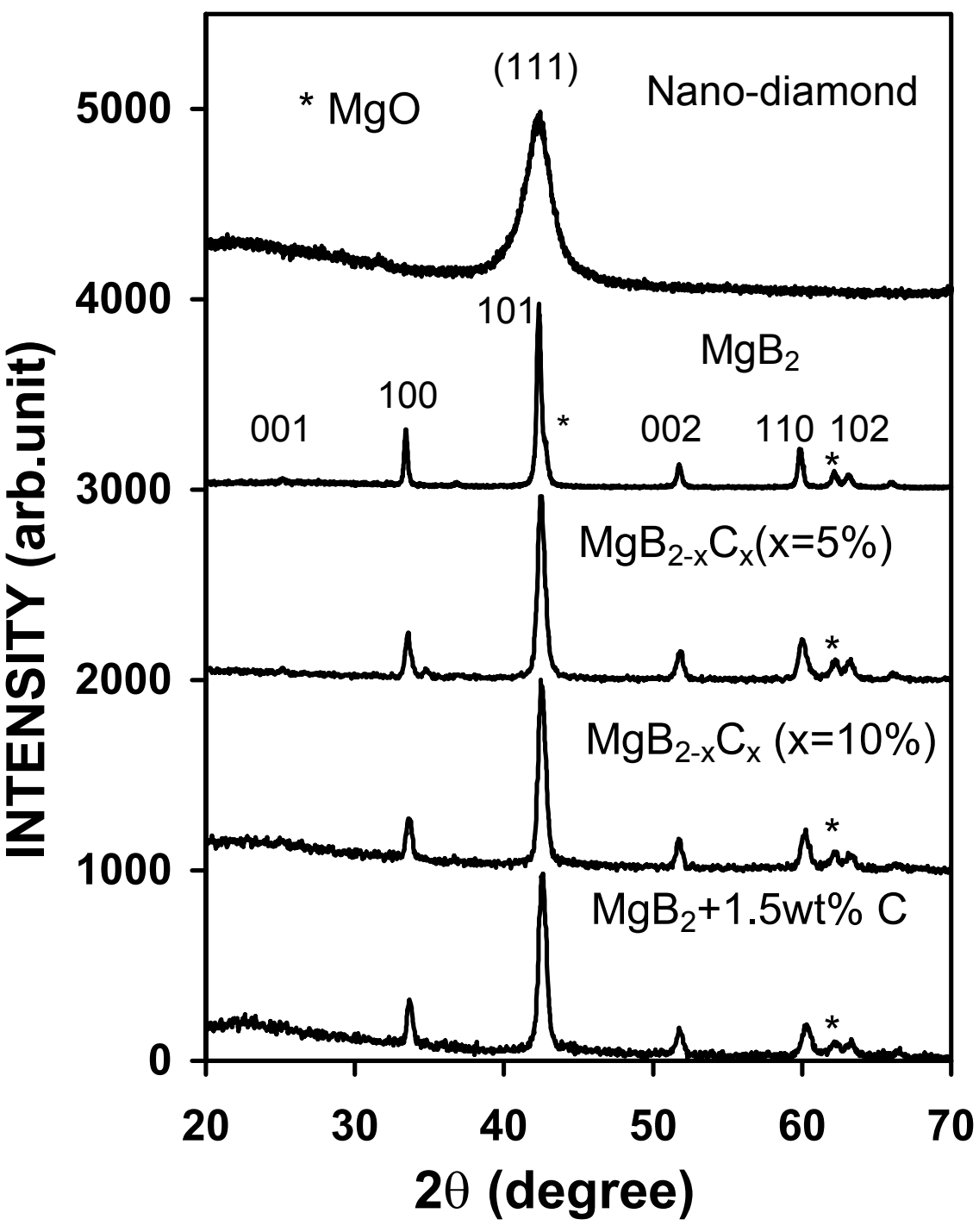

Fig.1

C.H. Cheng et al.

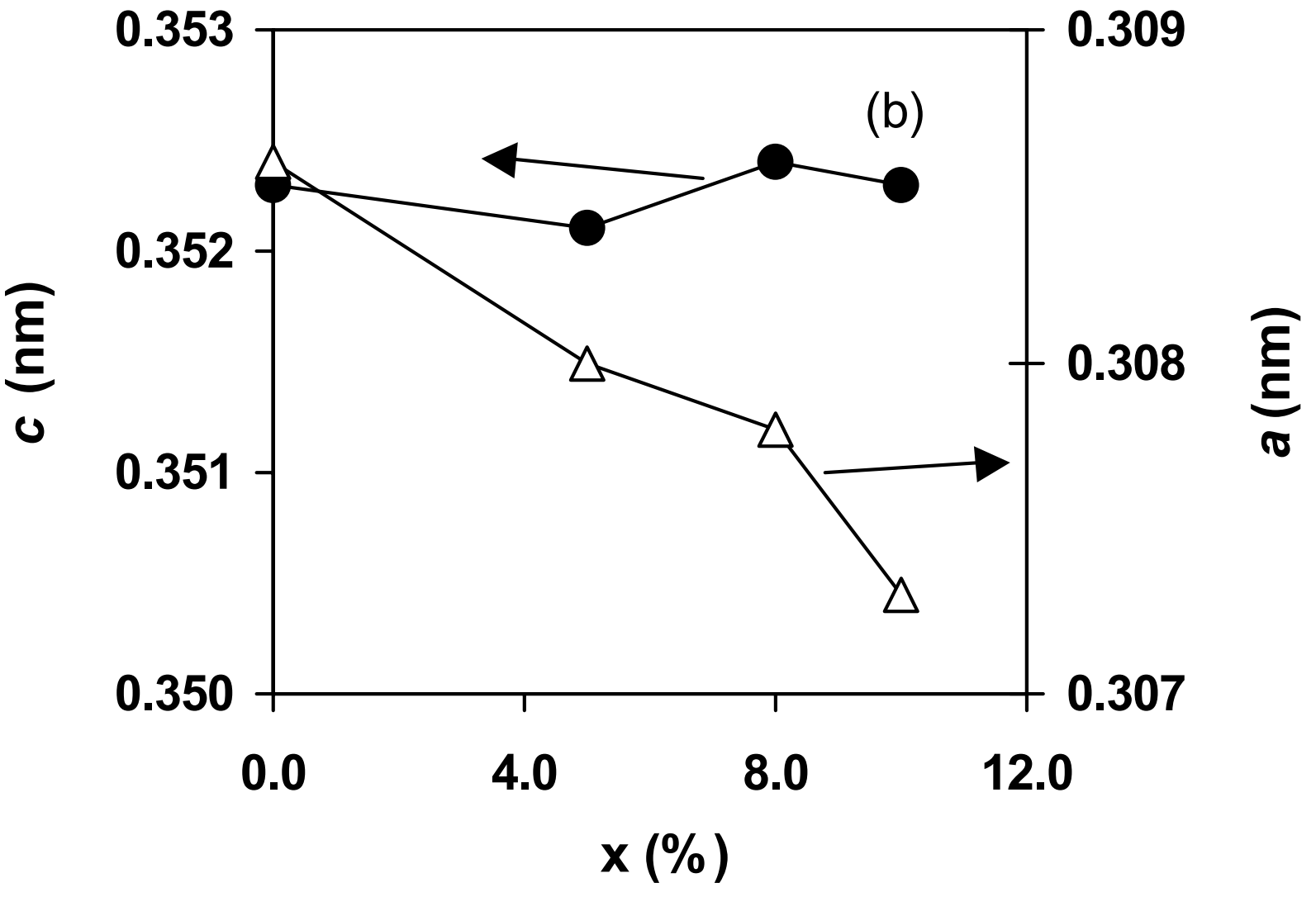

Fig.2

C.H. Cheng et al

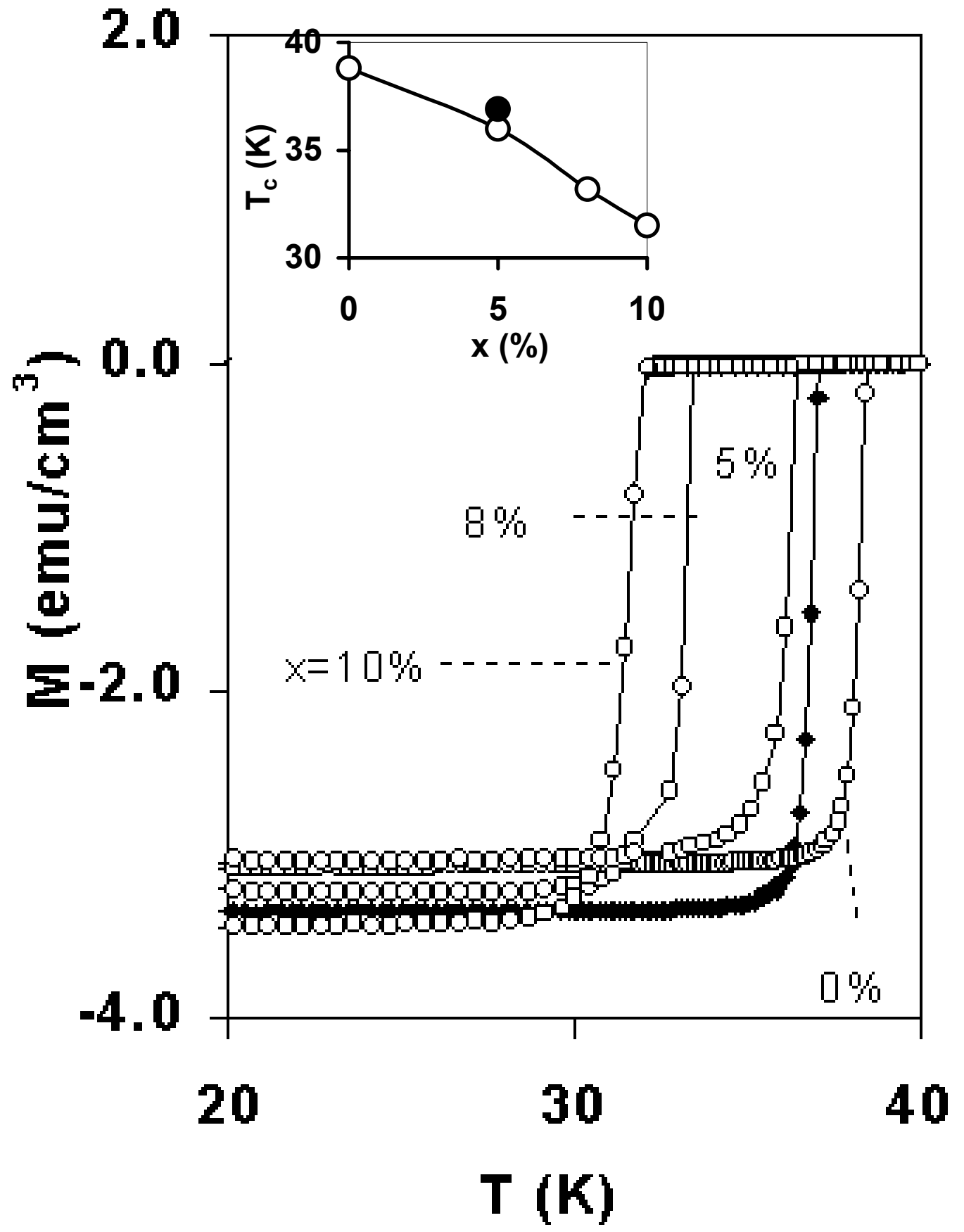

Fig. 3

C.H. Cheng et al

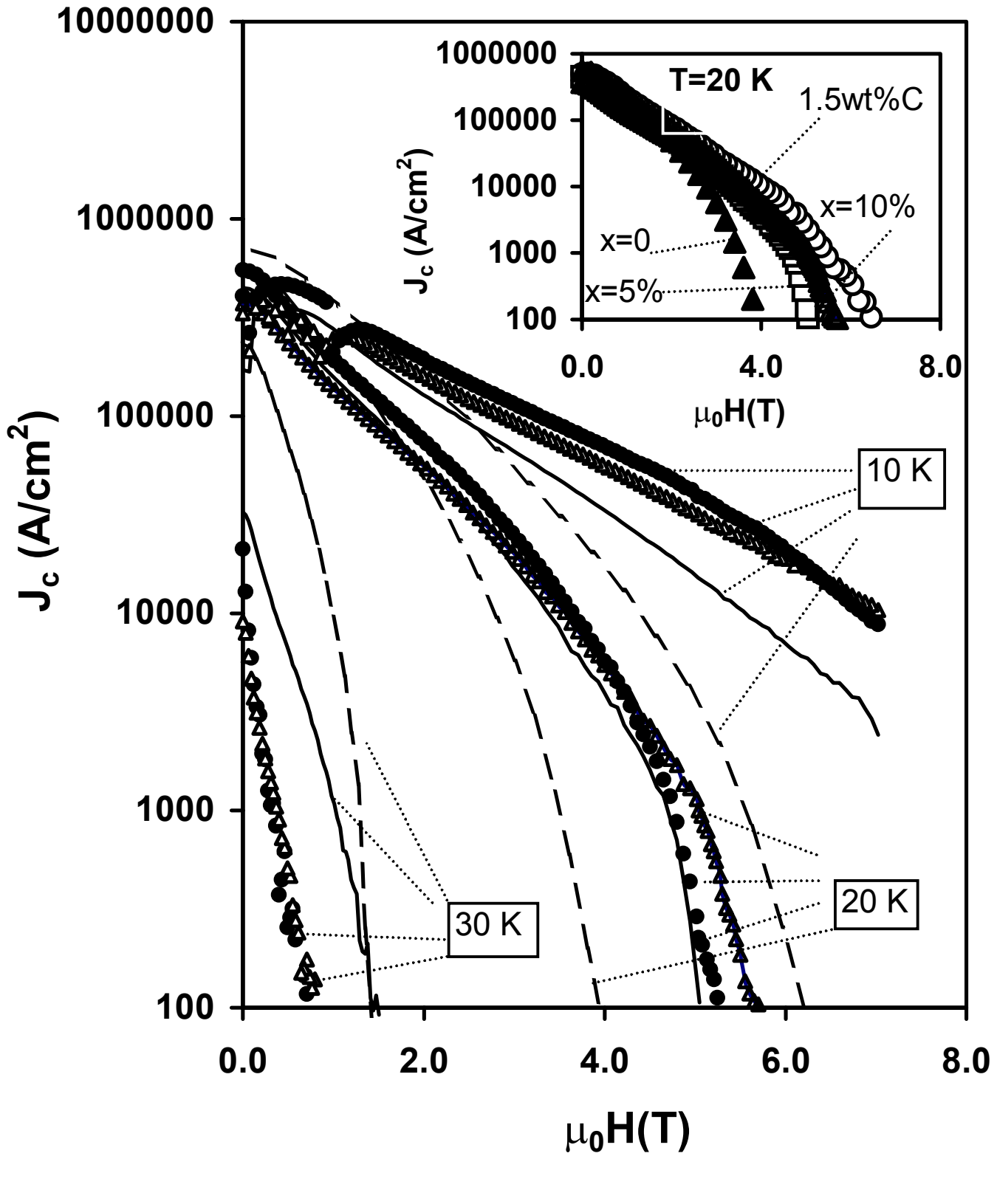

Fig.4
C.H. Cheng et al

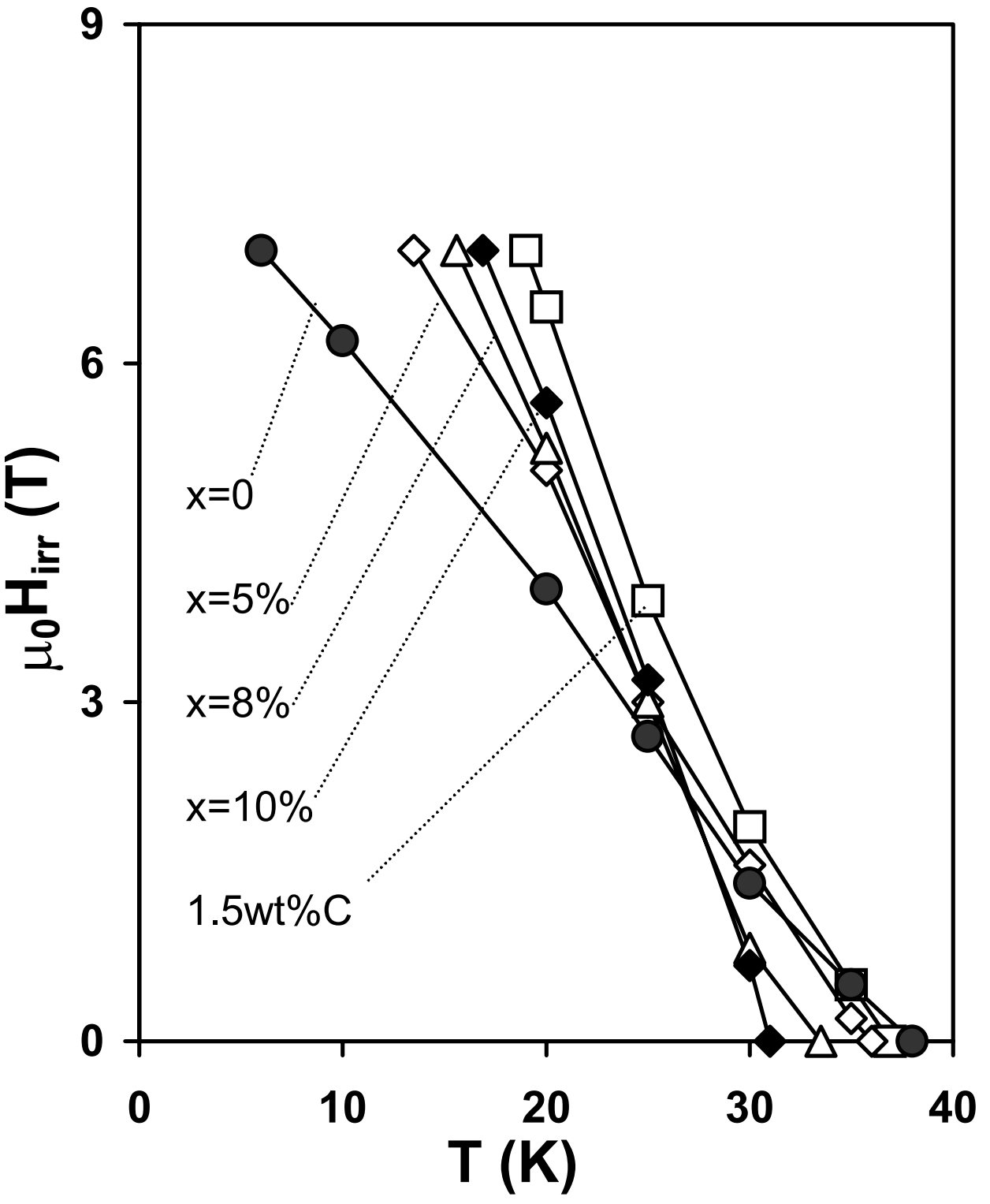

Fig.5

C.H. Cheng et al

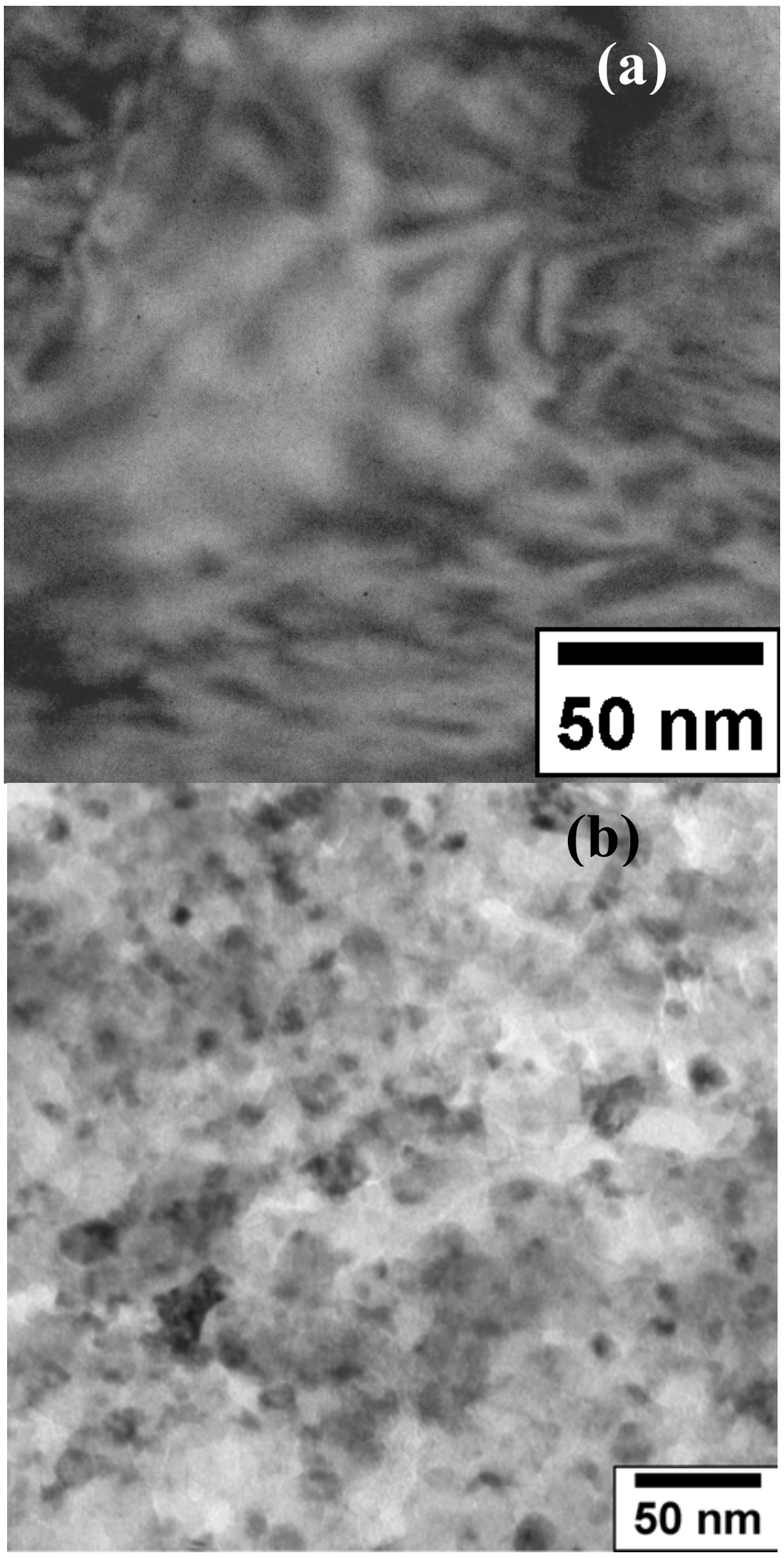

Fig. 6

C.H. Cheng et al